\documentclass[aps,pra,twocolumn,showpacs,letterpaper,superscriptaddress]{revtex4-1}

\usepackage[colorlinks=true, citecolor=blue, linkcolor=blue, urlcolor=blue]{hyperref}

\usepackage{graphicx,dcolumn,longtable,epsfig}
\usepackage[usenames]{color}
\usepackage{amssymb}
\usepackage{amsmath}
\usepackage{epstopdf}
\usepackage{bm}
\usepackage{footnote}
\usepackage{float}
\usepackage{subfigure}
\usepackage{color}
\usepackage{ulem}
\usepackage{lineno}
\usepackage[version=4]{mhchem}
\usepackage[T1]{fontenc}

\input epsf.tex

\def\bea{\begin{eqnarray}}
\def\eea{\end{eqnarray}}
\def\be{\begin{equation}}
\def\ee{\end{equation}}

\begin{document}
\title{
Supersolid phases of lattice dipoles tilted in three-dimensions}
\author{Jin Zhang}
\email{These authors contributed equally}
%\email{jin-zhang@uiowa.edu}
\affiliation{Department of Physics and Astronomy, University of Iowa, Iowa City, Iowa 52242, USA}
\author{Chao Zhang}
\email{These authors contributed equally}
\affiliation{State Key Laboratory of Precision Spectroscopy,
East China Normal University, Shanghai 200062, China}
\author{Jin Yang}
\email{jy9ug@virginia.edu}
\affiliation{Department of Physics, University of Virginia, Charlottesville, Virginia 22904, USA}
\author{Barbara Capogrosso-Sansone}
\affiliation{Department of Physics, Clark University, Worcester, Massachusetts 01610, USA}

\begin{abstract}
By means of quantum Monte Carlo simulations we study phase diagrams of dipolar bosons in a square optical lattice. The dipoles in the system are parallel to each other and their orientation can be fixed in any direction of the  three-dimensional space. Starting from experimentally tunable parameters like scattering length and dipolar interaction strength, we derive the parameters entering the effective Hamiltonian. 
Depending on the direction of the dipoles, various types of supersolids (e.g. checkerboard, stripe) and solids (checkerboard, stripe, diagonal stripe, and an incompressible phase) can be stabilized.
Remarkably, we find a cluster supersolid characterized by the formation of horizontal clusters of particles. These clusters order along a direction at an angle with the horizontal.  Moreover, we find what we call a grain-boundary superfluid. In this phase, regions with solid order are separated by extended defects --grain boundaries-- which support superfluidity. We also investigate the robustness of the stripe supersolid against thermal fluctuations. Finally, we comment on the experimental realization of the phases found.
\end{abstract}

\pacs{}
\maketitle

\section{Introduction}
\label{sec:sec1}
Supersolidity, a fascinating state of matter in which crystalline order and global phase coherence are simultaneously present, was originally predicted several decades ago~\cite{Gross,PenroseOnsager,THOULESS,JETP} and initially searched for in Helium systems~\cite{Kim:2004}. The quest for experimental realization of the supersolid phase has later focused on ultracold quantum gases as they offer a highly controllable platform where interactions can be finely tuned~\cite{RevModPhys.80.885}. The first observations of supersolidity in ultracold gases were made in atomic systems coupled with external light fields~\cite{Leonard:2017,Leonard1415,Li:2017}. In these setups, the density modulation is imposed by the external fields. More recently, experimentalists have exploited the anisotropic and long-range nature of dipolar interaction to demonstrate the existence of supersolid states of matter in ultracold dipolar gases~\cite{PfauPRX2019,ModugnoPRL2019,FerlainoPRX2019,Guo:2019,PhysRevLett.123.050402,Tanzi:2019, Tanzi1162,FerlainoNature2021, PhysRevLett.126.233401}. Here, the dipolar interaction is responsible for a spontaneous formation of droplets of gas organized in a crystalline structure (see also recent theoretical work in e.g. \cite{PhysRevResearch.3.033125,Kora:2019aa}).

Supersolid structures have also been theoretically predicted in dipolar gases trapped in optical lattices~\cite{Danshita:2009cp,CapogrossoSansone:2010em,PhysRevA.83.013627,CapogrossoSansone:2011eq,Bandyopadhyay:2019ew,Zhang:2018ew,Kraus:2020es,Giovannastagger,PhysRevA.103.043333, Wu:2020kf, PhysRevLett.84.1599}. 
As optical lattices already impose a crystalline structure, solid order in these systems is realized when a discrete symmetry is also broken as particles arrange themselves in a crystalline structure different than the one of the underlining optical lattice, e.g. checkerboard, stripe patterns. 
 While supersolidity in ultracold atoms trapped in optical lattices has yet to be experimentally observed, a recent experiment~\cite{Baier:2016ga} has paved the way to investigate this elusive phase with dipolar lattice bosons.

Motivated by these recent experimental breakthroughs, here, we study under which experimental conditions supersolid phases can be realized with lattice dipolar bosons tilted in three dimensions (3D). We use quantum Monte Carlo simulations based on the worm algorithm~\cite{Prokofev:1998gz} to study quantum phases stabilized by the extended Bose-Hubbard model in a square lattice. The model describes a system of soft-core dipolar bosons with dipoles   parallel to each other. The polarization axis can be fixed in any direction of the  three-dimensional space. We notice that we calculate the
parameters entering the effective Bose-Hubbard model, i.e. the onsite
interaction, long-range interaction strength, and density-induced hopping, from the parameters that can be tuned experimentally, such as scattering length, dipolar interaction strength, and optical lattice potential depth.
This paper is organized as follows: In Sec.~\ref{sec:sec2} we introduce the Hamiltonian of the system and the relative parameters that can be controlled in experiments. In Sec.~\ref{sec:sec3} we discuss various phases and the corresponding order parameters. In Sec.~\ref{sec:sec4} we present the phase diagrams with different dipole orientations and other experimentally controllable conditions. 
In Sec.~\ref{sec:sec5} we discuss the experimental realization. We conclude the article in Sec.~\ref{sec:sec6}.

\section{Hamiltonian} 
\label{sec:sec2}

\begin{figure}[h]
\includegraphics[trim=11cm 0cm 12cm 0cm, clip=true, width=0.48\textwidth]{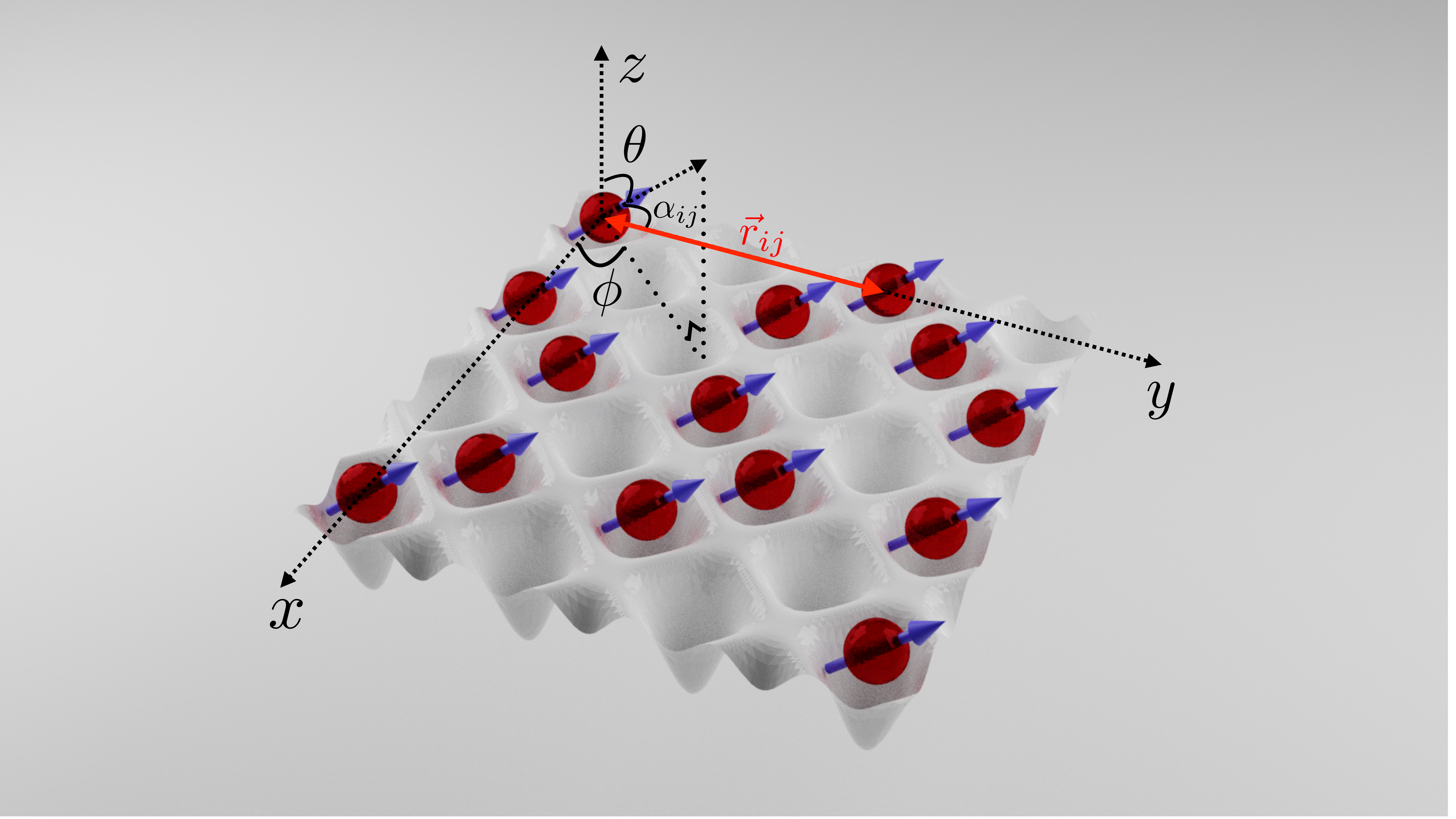}
\caption{Schematic representation of the system. Dipoles are trapped in a two-dimensional optical lattice and are aligned parallel to each other along the direction of polarization, determined by an electric/magnetic field. $\theta$ is the polar angle between polarization axis and $z$ direction, $\phi$ is the azimuthal angle. $\vec{r}_{ij}$ is the relative position between site $i$ and $j$. $\alpha_{ij}$ is the angle between polarization axis and $\vec{r}_{ij}$.}
\label{FIG6}
\end{figure}

%%%%%%%%%%%%%%%%%%%%%%%%%%%%%%%
We study a two-dimensional cold-atom system of dipolar bosons in a square optical lattice as shown in Fig.~\ref{FIG6}. The external potential that creates the lattice is given by $V_{\text{ext.}}(x,y,z) = V_0 \left[ \cos^2(k_L x) + \cos^2(k_L y) \right] + m \Omega^2_z z^2 / 2$, where $V_0$ denotes the depth of the 2D optical lattice, $k_L = 2\pi/\lambda = \pi / a$ is the lattice momentum, $a=\lambda/2$ is the lattice spacing, $m$ is the atomic mass, and $\Omega_z$ is the angular frequency of the harmonic trap in $z$ direction. The lattice depth $s$ is expressed in units of recoil energy, $s = V_0 / E_R$, with $E_R=\frac{\hbar^2k_L^2}{2m}$. The $z$ dependence of the wavefunction is a Gaussian function that in lattice coordinates, $\mathbf{r} \rightarrow \mathbf{r}/a$,   reads as $\sim\exp(-\pi^2 \kappa z^2 / 2)$, where $\kappa = \hbar \Omega_z/2E_R$ is the flattening constant \cite{Sowinski:2012kl} characterizing the width of the 2D sheet for our system. The dipole moments are allowed to rotate in three-dimensional space and are characterized by the polar angle $\theta$ between the dipole moment and the $z$ axis and the azimuthal angle $\phi$ (see Fig.~\ref{FIG6}).

In second quantization and Wannier basis \cite{Kohn:1959bn}, one can obtain the 2D extended Bose-Hubbard (EBH) model for the lowest Bloch band,
\begin{align}
\nonumber
H=&-t \sum_{\langle\mathbf{i}, \mathbf{j}\rangle} a_{\mathbf{i}}^{\dagger} a_{\mathbf{j}}+\frac{U}{2} \sum_{\mathbf{i}} n_{\mathbf{i}}\left(n_{\mathbf{i}}-1\right)+\frac{1}{2} \sum_{\mathbf{i}, \mathbf{j}} V_{\mathbf{i}, \mathbf{j}} n_{\mathbf{i}} n_{\mathbf{j}} \\
&-\sum_{\langle\mathbf{i}, \mathbf{j}\rangle} T_{\mathbf{i}, \mathbf{j}} a_{\mathbf{i}}^{\dagger}\left(n_{\mathbf{i}}+n_{\mathbf{j}}\right) a_{\mathbf{j}}-\mu \sum_{\mathbf{i}} n_{\mathbf{i}} ,
\label{Eq1}
\end{align}
where $a_{\mathbf{i}}^\dagger$ ($a_{\mathbf{i}}$) are bosonic creation (annihilation) operators satisfying the bosonic commutation relations $[a_{\mathbf{i}}, a_{\mathbf{j}}^\dagger] = \delta_{\mathbf{i}\mathbf{j}}$, $n_{\mathbf{i}}=a_{\mathbf{i}}^{\dagger}a_{\mathbf{i}}$ is the particle number operator, $t$ is the amplitude of nearest-neighbor tunneling, $U$ is the onsite interaction. $V_{\mathbf{i}\mathbf{j}}$ is the off-site interaction between atoms on site $\mathbf{i}$ and site $\mathbf{j}$. $T_{\mathbf{i}\mathbf{j}}$ is amplitude of the density-induced hopping, and $\mu$ is the chemical potential which we vary in our simulations to achieve a specific filling. Here $\langle \cdots \rangle$ denotes nearest-neighboring (NN) sites. We consider all off-site interaction terms within $|\mathbf{i} - \mathbf{j}| \le 5$ to include long-range dipolar interactions. We set the lattice depth $s = 10$, which gives the nearest-neighbor hopping amplitude $t = 0.0192 E_R$. 

The interaction between dipolar bosonic fields residing at $\mathbf{r}$ and $\mathbf{r}'$ contains contact interaction $V_c$ and dipole-dipole interaction $V_{dd}$,
\begin{eqnarray}
\nonumber V(\mathbf{r}-\mathbf{r}^{\prime}) &=& V_c(\mathbf{r}-\mathbf{r}^{\prime}) + V_{dd}(\mathbf{r}-\mathbf{r}^{\prime}) \\ 
&=& g \delta(\mathbf{r}-\mathbf{r}^{\prime}) + \gamma \frac{1-3\cos^2(\alpha)}{|\mathbf{r}-\mathbf{r}^{\prime}|^3},
\end{eqnarray}
where $g = 8 a_s / (\pi a)$, $a_s$ is the $s$-wave scattering length, and $\gamma = m \mu_e^2 / (2\pi^3 \epsilon_0 \hbar^2 a)$ for electric dipolar interactions, or $\mu_0 \mu_m^2 m / (2\pi^3\hbar^2 a)$ for magnetic dipolar interactions. $\mu_e$ ($\mu_m$) is the electric (magnetic) dipole moment of the bosons, $\epsilon_0$ ($\mu_0$) is the vacuum permittivity (permeability), and $\alpha$ is the angle between the dipole moments and the relative position of the two bosons $\mathbf{r}-\mathbf{r}'$. Thus, the parameters in the effective Hamiltonian~\eqref{Eq1} have contributions from both contact and dipolar interactions: $U = U^c + U^{d d}$, $T_{\mathbf{i},\mathbf{j}} = T_{\mathbf{i},\mathbf{j}}^c - T_{\mathbf{i},\mathbf{j}}^{d d}$, and $V_{\mathbf{i},\mathbf{j}} = V_{\mathbf{i},\mathbf{j}}^c + V_{\mathbf{i},\mathbf{j}}^{d d}$ \cite{Dutta:2015fe}. For $\kappa = 10$ and $a_s / a = 0.014$, which is typical in current experiments, the values of the contact part are $U^c / t = 30.5, V^c_{\langle \mathbf{i}, \mathbf{j} \rangle} / t = 0.006, T^c / t = 0.104$. 
The contact part of off-site interaction is negligible, so $V_{\mathbf{i},\mathbf{j}}$ is dominated by the dipolar contribution. The contact contribution is proportional to $g$ and $\sqrt{\kappa}$ (from the $z$-dependent Gaussian part of the wavefunction) \cite{PhysRevA.103.043333}, so the values of $U^c, V^c, T^c$ for all $g$ and $\kappa$ are easy to calculate. We set $\kappa = 10$ in all our calculations. For each set of $(U, \theta, \phi)$, we vary the value of $\gamma$ and study the phase diagram of the system.

\begin{figure}[h]
\includegraphics[width=0.5\textwidth]{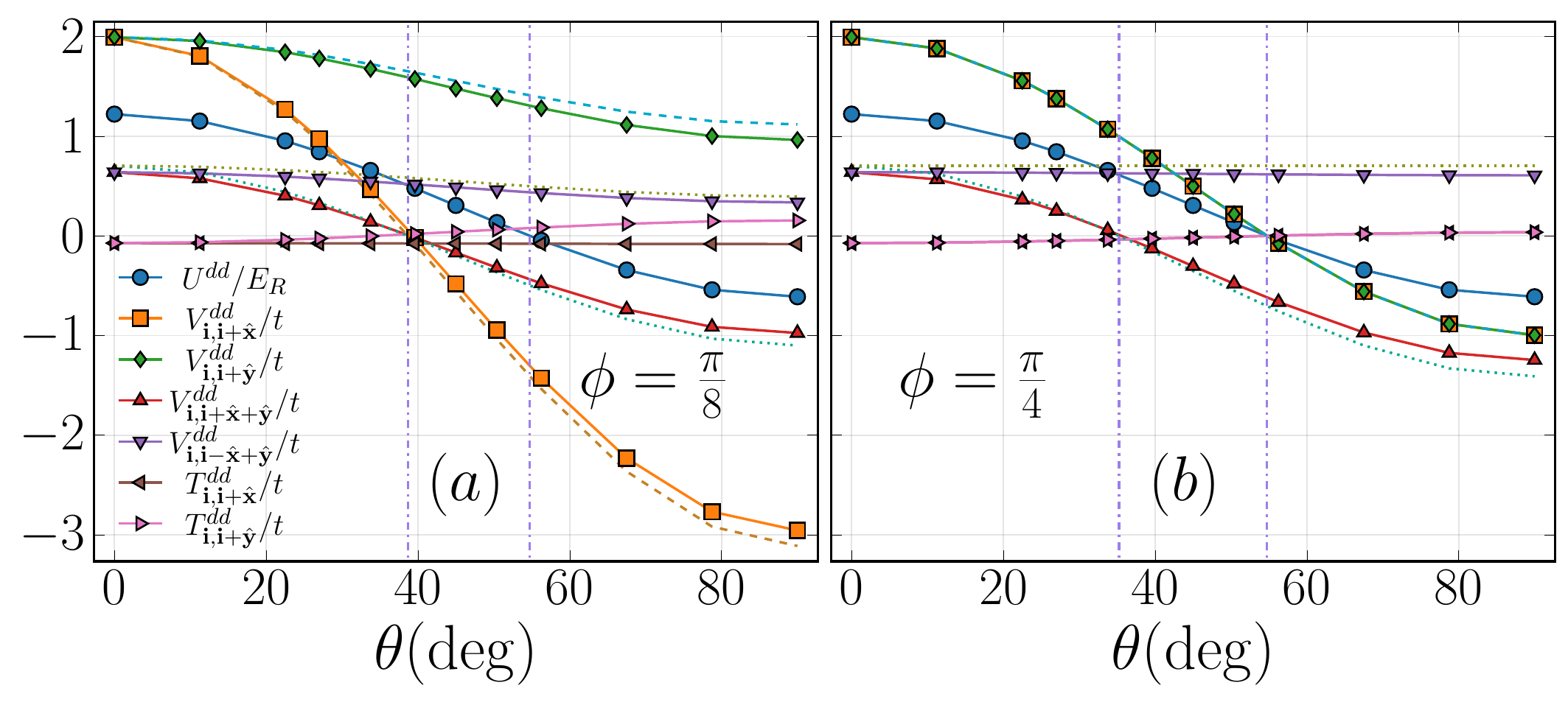}
\caption{Dipolar contribution to Hamiltonian parameters as a function of polar angle $\theta$ at $\gamma = 1/\pi^3, s = 10, \kappa = 10$, (a) $\phi=\pi/8$, and (b) $\phi=\pi/4$. The two vertical dashdotted lines locate angles $\theta = 38.7^{\circ},54.7^{\circ}$ for $\phi=\pi/8$, and $\theta = 35.3^{\circ},54.7^{\circ}$ for $\phi = \pi/4$. The dashed line by $V_{\mathbf{i}, \mathbf{i}+\hat{\mathbf{x}}}^{d d} / t$ is $V[1-3\sin^2(\theta)\cos^2(\phi)]$, and the dashed line by $V_{\mathbf{i}, \mathbf{i}+\hat{\mathbf{y}}}^{d d} / t$ is $V[1-3\sin^2(\theta)\sin^2(\phi)]$. The dotted line by $V^{d d}_{\mathbf{i},\mathbf{i}+\hat{\mathbf{x}}+\hat{\mathbf{y}}}/t$ is $V\left(1-3\sin^2(\theta)\cos^2(\phi-45^{\circ})\right) / (\sqrt{2})^3$, and the dotted line by $V^{d d}_{\mathbf{i},\mathbf{i}-\hat{\mathbf{x}}+\hat{\mathbf{y}}}/t$ is $V\left(1-3\sin^2(\theta)\cos^2(\phi-135^{\circ})\right) / (\sqrt{2})^3$. $V$ is the nearest-neighbor interaction at $\theta = \phi = 0$. The dashed lines in (b) are on top of the calculated lines.}
\label{fig:hamparamsvsthetaphi}
\end{figure}

\begin{figure*}
\includegraphics[width=\textwidth]{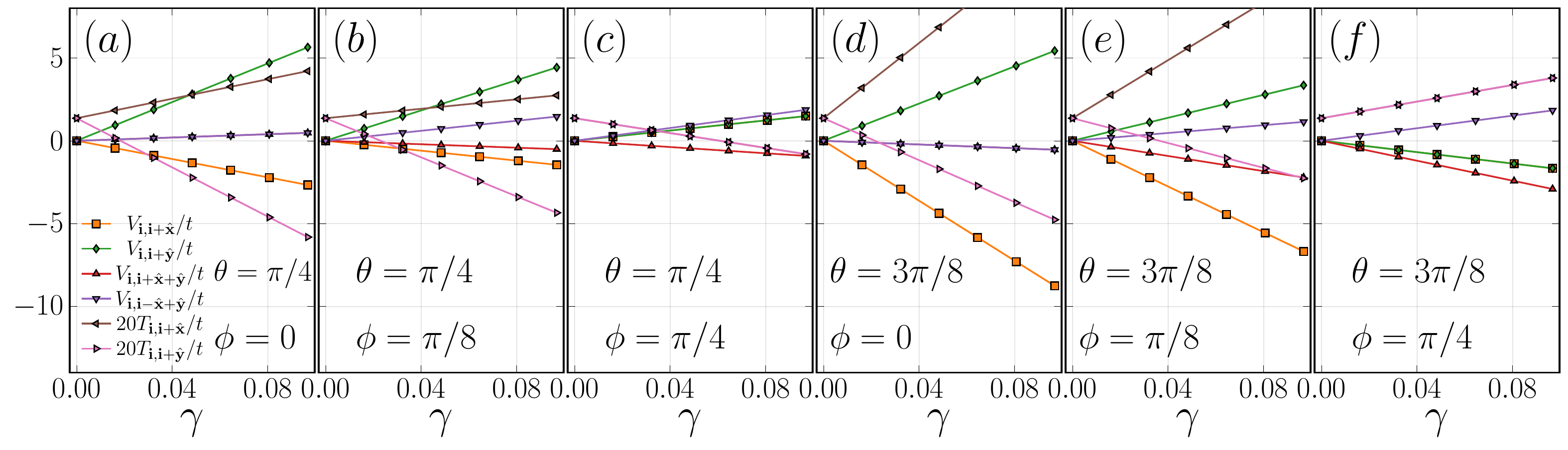}
\caption{Hamiltonian parameters as a function of $\gamma$ for different values of $\theta$ and $\phi$. The total onsite interaction $U/t = 20$ and $\kappa = 10$. For $\phi = \pi/4$, the data points for $V_{\mathbf{i}, \mathbf{i}+\hat{\mathbf{x}}}$ and $V_{\mathbf{i}, \mathbf{i}+\hat{\mathbf{y}}}$ ($T_{\mathbf{i}, \mathbf{i}+\hat{\mathbf{x}}}$ and $T_{\mathbf{i}, \mathbf{i}+\hat{\mathbf{y}}}$) are on top of each other.}
\label{fig:hamparamsvsgamma}
\end{figure*}

In Fig.~\ref{fig:hamparamsvsthetaphi}, we present the dipolar contribution to the Hamiltonian parameters as functions of $\theta$ for $\gamma = 1/\pi^3$, $\kappa = 10$ and $\phi = \pi/8, \pi/4$. Notice that the onsite interaction is in units of $E_R$, while others are in units of $t$. The onsite interaction does not depend on the azimuthal angle. For an ideal 2D system, the $z$-dependent Gaussian part of the wavefunction contributes a factor of $\sqrt{\kappa}$ to $U^{dd}$. We observe that $U^{d d} = 0$ at $\theta = \sin^{-1}(\sqrt{2/3})$, which is approximately $\theta\sim54.7^{\circ}$. Using the effective 2D interaction in Eq. (A6) from Ref.~\cite{PhysRevA.103.043333}, it is easy to prove that $U^{dd} = 0$ at $\sin^2(\theta) = 2/3$, independent of other experimental parameters. For off-site interactions, the values are all close to the approximation
\begin{eqnarray}
\label{eq:vapprox}
V^{dd}_{\mathbf{i},\mathbf{j}} \approx \frac{V^{dd}(1-3\cos^2(\alpha))}{|\mathbf{i}-\mathbf{j}|^3},
\end{eqnarray}
where $V^{dd}$ is the nearest-neighbor dipolar part of the interaction at $\theta = 0$. This approximation is valid for deep enough lattice potential. In the $x$ direction, $\cos(\alpha) = \sin(\theta)\cos(\phi)$, $V^{dd}_{\mathbf{i},\mathbf{i}+\hat{\mathbf{x}}} \approx 0$ at $\theta \approx 38.7^{\circ}, 54.7^{\circ}$ for $\phi = \pi/8, \pi/4$ respectively. In the $y$ direction, $\cos(\alpha) = \sin(\theta)\sin(\phi)$, $V^{dd}_{\mathbf{i},\mathbf{i}+\hat{\mathbf{y}}}$ is always positive for $\phi = \pi/8$, and $V^{dd}_{\mathbf{i},\mathbf{i}+\hat{\mathbf{y}}} \approx 0$ at $\theta \approx 54.7^{\circ}$ for $\phi = \pi/4$. The value of $V^{dd}_{\mathbf{i},\mathbf{i}+\hat{\mathbf{x}}}$ and that of $V^{dd}_{\mathbf{i},\mathbf{i}+\hat{\mathbf{y}}}$ should be the same at $\phi = \pi/4$. In $\hat{\mathbf{x}} + \hat{\mathbf{y}}$ direction, $\cos(\alpha) = \sin(\theta)\cos(\phi-\pi/4)$, $V^{dd}_{\mathbf{i},\mathbf{i}+\hat{\mathbf{x}}+\hat{\mathbf{y}}} \approx 0$ at $\theta \approx 38.7^{\circ}, 35.3^{\circ}$ for $\phi = \pi/8, \pi/4$ respectively. In $-\hat{\mathbf{x}} + \hat{\mathbf{y}}$ direction, $\cos(\alpha) = \sin(\theta)\cos(\phi-3\pi/4)$, $V^{dd}_{\mathbf{i},\mathbf{i}-\hat{\mathbf{x}}+\hat{\mathbf{y}}}$ is always positive for $\phi = \pi/8$, and it is independent of $\theta$ for $\phi=\pi/4$. These expected behaviors are all confirmed in Fig.~\ref{fig:hamparamsvsthetaphi}. Furthermore, the approximation in Eq.~\eqref{eq:vapprox} becomes exact for nearest-neighbor interactions in both directions at $\phi = \pi/4$. The off-site interactions $V^{dd}_{\mathbf{i},\mathbf{j}}$ has little dependence on the value of $\kappa$ for $\kappa \gtrsim 6$ \cite{PhysRevA.103.043333}, indicating that the system can be approximated as a 2D one. 
One can see that $T_{\mathbf{i},\mathbf{i+\hat{\mathbf{x}}}}^{dd}$ and $T_{\mathbf{i},\mathbf{i+\hat{\mathbf{y}}}}^{dd}$ are close to zero and increase slowly with increasing $\theta$ for both values of $\phi$ and $\kappa = 10$. As a consequence, at low filling, we do not expect significant changes in the phase diagrams compared to the case with no density-induced hopping. Notice that for $\theta = 54.7^{\circ}, \phi = \pi/4$, the dipolar part of the onsite interaction, the off-site interactions in both $\hat{\mathbf{x}}$ and $\hat{\mathbf{y}}$ directions, and the density-induced hopping are all zero.

In our calculations, we fix the value of the total onsite interaction. Then, for a given dipole orientation, other Hamiltonian parameters only depend on $\gamma$.  Figure~\ref{fig:hamparamsvsgamma} depicts the dependence of the Hamiltonian parameters on $\gamma$ for $\kappa = 10$, $\theta = \pi/4, 3\pi/8$, and $\phi = 0, \pi/8, \pi/4$. Notice that we multiply $T/t$ by $20$ for a better view. As the contact part of the NN interaction ($V^c/t = 0.006$) is close to zero, the total off-site interactions are dominated by the dipolar part. Comparing Figs.~\ref{fig:hamparamsvsgamma}(a), (b), and (c), we notice that the magnitude of the NN interactions becomes smaller as we increase the azimuthal angle $\phi$ for fixed $\theta = \pi/4$, consistent with the results in Fig.~\ref{fig:hamparamsvsthetaphi}. Thus we expect, for larger $\phi$, the superfluid phase can persist at larger values of $\gamma$. At small $\phi$, the NN interaction in $y$ direction is repulsive and stronger than the attractive interaction in $x$ direction, thus particles tend to populate rows separated by empty ones. The NN interactions become identical and repulsive at $\phi=\pi/4$, smaller than the NNN interaction in $-\hat{\mathbf{x}}+\hat{\mathbf{y}}$ direction. Since the NNN interaction in $\hat{\mathbf{x}}+\hat{\mathbf{y}}$ direction is attractive, the particles tend to populate in diagonal lines in $\hat{\mathbf{x}}+\hat{\mathbf{y}}$ direction. Figures~\ref{fig:hamparamsvsgamma}(d), (e), and (f) show that at $\theta = 3\pi/8$, the magnitude of the NN interactions also decrease as we increase $\phi$. 
But at this $\theta$, the attractive parts of the off-site interaction are stronger than the repulsive parts. In the point of view of the mean-field approximation, the off-site interactions in every direction are effective linear potentials depending on local densities, and the total potential for each site is negative, thus the system should stabilize at a finite density. We will see that the competition between attractive and repulsive interaction tends to destabilize solid phases at filling smaller than one. Low densities turn out to be unstable and no quantum phases can be stabilized while, at larger densities, particles tend to occupy every site of the lattice and a SF phase is stabilized. We expect solid phases to be stable at densities $n>1$, not considered here. 
Finally, the density-induced hoppings at $\theta = 3\pi/8$ is larger than those at $\theta = \pi/4$, so we can expect a stronger superfluid response. In the following, we discuss the phase diagrams for these cases.

%%%%%%%%%%%%%%%%%%%%%%%%%%%%%%%

\section{Quantum phases and order parameters}
\label{sec:sec3}

In this section, we present some of the phases stabilized by Eq.~\eqref{Eq1} and the corresponding order parameters. Fig.~\ref{Table1} shows order parameters for the superfluid (SF) phase, checkerboard solid (CB) phase, checkerboard supersolid (CBSS) phase, stripe solid (StrS) phase, stripe supersolid (StrSS) phase, and diagonal stripe solid phase (DiagStrS).
Each phase corresponds to a unique combination of the order parameters. In order to characterize these quantum phases we need the following order parameters: superfluid density $\rho_s$, structure factor $S(\pi, \pi)$, $S(0, \pi)$, and $S(\pi/2, -\pi/2)$. Notice that we have found other quantum phases that are not captured by these order parameters (see next Section for details).

\begin{table}[h]
\includegraphics[trim=2cm 16cm 12cm 2cm, clip=true, width=0.6\textwidth]{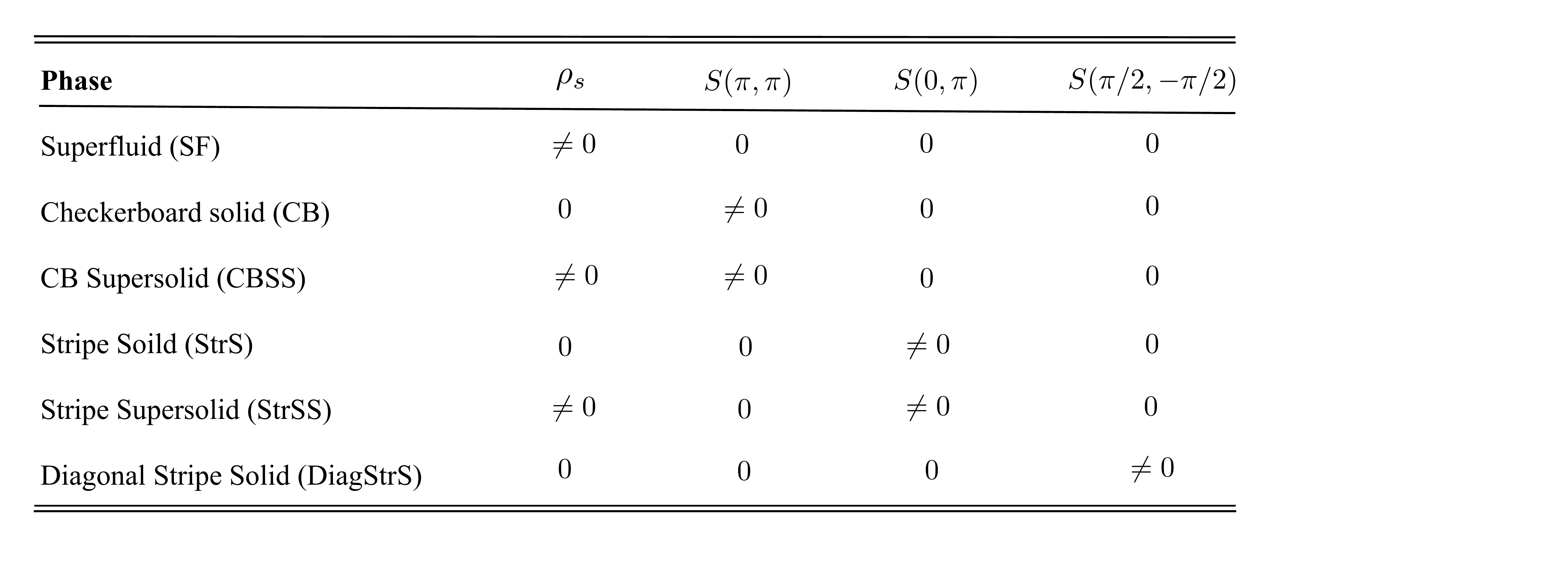}
\caption{ Quantum phases and the corresponding order parameters: superfluid density $\rho_s$, structure factor $S(\pi, \pi)$, $S(0,\pi)$, and $S(\pi/2, - \pi/2)$.
}
\label{Table1}
\end{table}

The superfluid density is calculated in terms of the winding number~\cite{Winding}: $\rho_s=\langle \mathbf{W}^2 \rangle /DL^{D-2}\beta$, where $\langle\mathbf{W}^2\rangle=\sum_{i=1}^D\langle W_i^2\rangle$ is the expectation value of winding number square, $D$ is the dimension of the system and here $D=2$, $L$ is the linear system size, and $\beta$ is the inverse temperature. The structure factor characterizes diagonal long-range order and is defined as: $S(\mathbf{k})=\sum_{\mathbf{r},\mathbf{r'}} \exp{[i \mathbf{k}\cdot (\mathbf{r}-\mathbf{r'})]\langle n_{\mathbf{r}}n_{\mathbf{r'}}\rangle}/N$, where $N$ is the particle number. $\mathbf{k}$ is the reciprocal lattice vector. We
use $\mathbf{k}=(\pi, \pi)$, $\mathbf{k}=(0,\pi)$, and $\mathbf{k}=(\pi/2, -\pi/2)$ to identify the CB, StrS, and DiagStrS, respectively.  Another quantity we monitor is compressibility  defined as $\frac{\beta\Delta N^2}{L^2}$, where $\Delta N^2=\langle(N-\langle N\rangle)^2\rangle$.
The compressibility is finite for compressible phases and zero (in the thermodynamic limit) for incompressible phases.

\section{Ground-state phase diagrams}
\label{sec:sec4}

\begin{figure*}[th]
\centering
\includegraphics[trim=0.5cm 0cm 0cm 0cm, clip=true, width=1.0\textwidth]{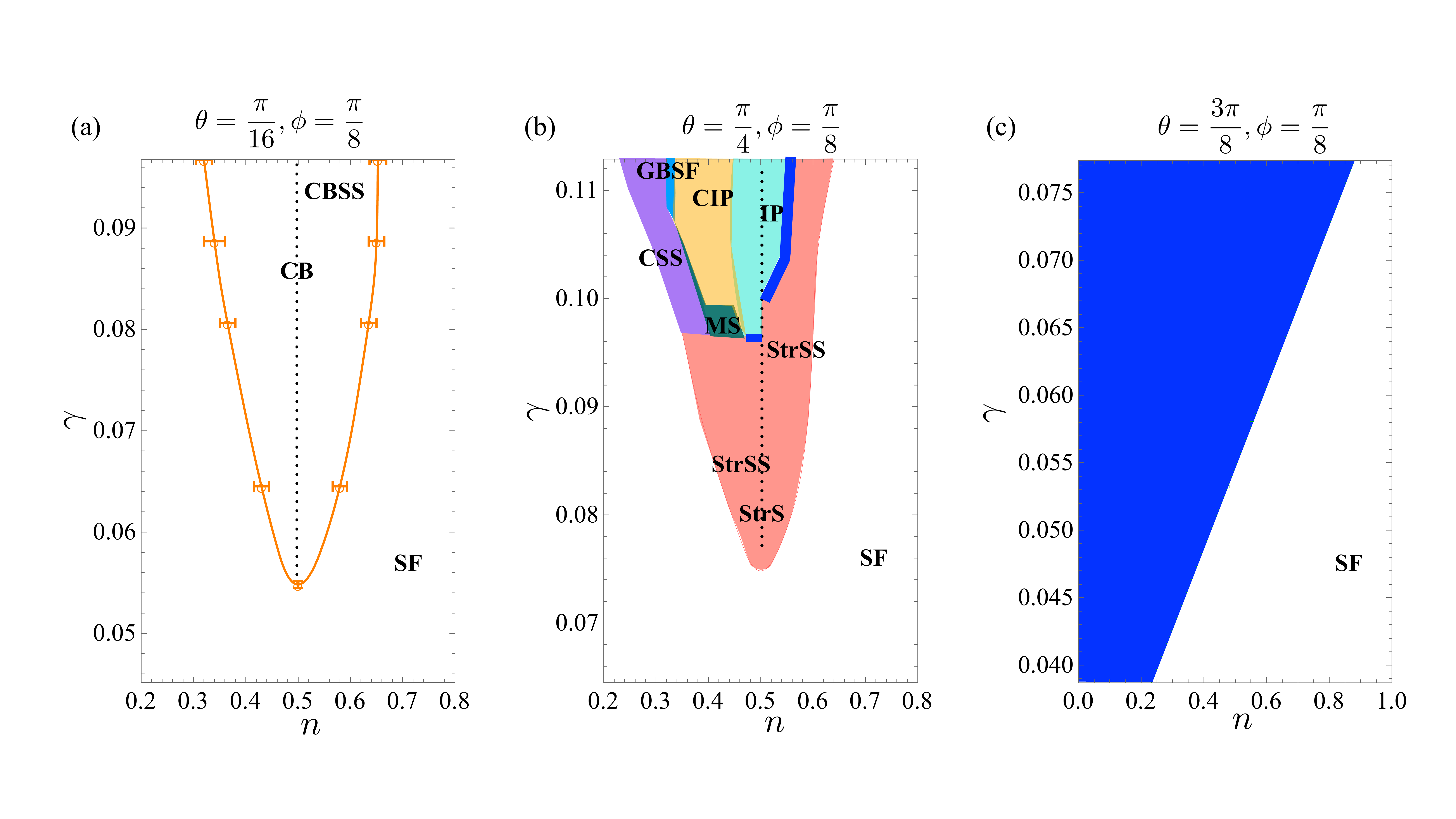}
\caption{Ground-state phase diagrams for $\phi=\frac{\pi}{8}$, $\theta=\frac{\pi}{16}$(a), $\theta=\frac{\pi}{4}$(b), and $\theta=\frac{3\pi}{8}$(c). The $x$-axis is the filling factor $n$ and the $y$-axis is the dipolar interaction strength $\gamma$. For polar angles $\theta\lesssim \frac{\pi}{6}$, the solid phase stabilized at half filling is a  checkerboard solid (CB) and the supersolid phase is a checkerboard supersolid (CBSS); for $\theta\gtrsim \frac{\pi}{6}$, the half filling solid phase is a stripe solid (StrS) and the supersolid phase around half filling is a stripe supersolid (StrSS). In (b), CSS stands for cluster supersolid, IP stands for the incompressible ground states stabilized at rational filling factors, CIP is a cluster incompressible phase (see text for more details), GBSF is a grain-boundary superfluid (see text for more details), and MS is a metastable region. Dark blue regions in (b) and (c) correspond to first-order phase transitions. Dotted lines at filling factor $n=0.5$ represent solid phases CB or StrS.}
\label{FIG1}
\end{figure*}

Throughout this section, we fix $U/t=20$ and flattening constant $\kappa=10$, and study under which experimental parameters supersolids and other phases are stabilized for filling factors $n<1$. The results presented are an extended investigation of what was discussed in~\cite{PhysRevA.103.043333}, where dipoles are tilted within the $x$-$z$ plane, that is, the azimuthal angle is fixed at $\phi=0$. In this work, dipoles are parallel to each other and the polarization axis can be fixed in any direction in the 3D space. Due to the reflection symmetry along the diagonal of the square lattice, we only consider $\phi\leqslant\pi/4$. We investigate the phase diagrams at $\phi=\pi/8$ and $\pi/4$ for three values of the polar angle $\theta=\pi/16$, $\pi/4$, $3\pi/8$. The polar angles are chosen so that we can make comparisons with results presented in~\cite{PhysRevA.103.043333}, where $\phi=0$.
System sizes $L=20$, 32, and 40 are used to get the transition points on phase diagrams, and the inverse temperature is set to $\beta=L$. 

\begin{figure*}[th]
\centering
\includegraphics[trim=0.5cm 0cm 5cm 0cm, clip=true, width=0.98\textwidth]{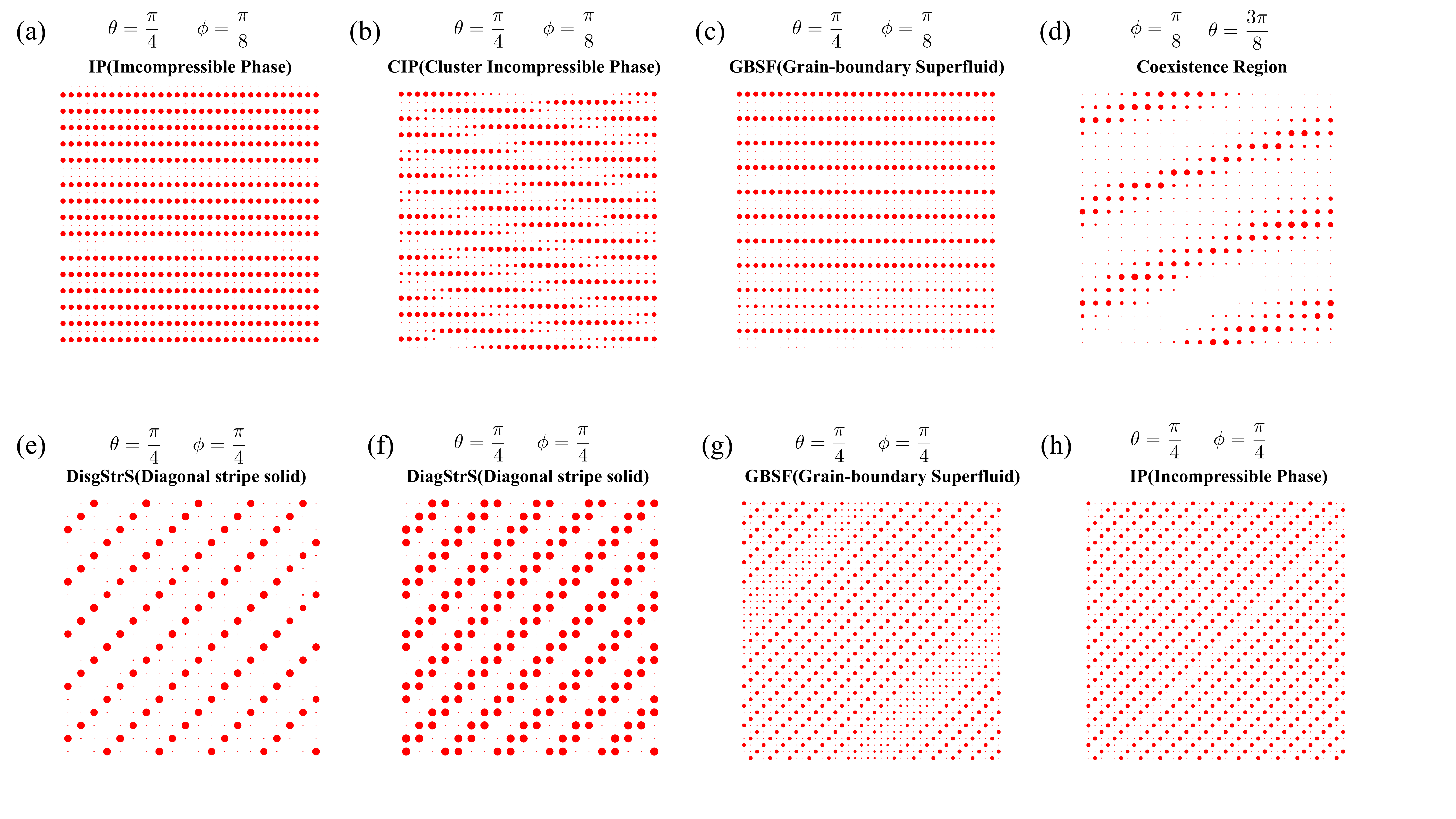}
\caption{ Density maps for various phases. Each circle corresponds to a different site and its radius is proportional to the local density. (a-c) $\theta=\frac{\pi}{4}$, $\phi=\frac{\pi}{8}$, $\gamma=0.1096$, and $L=32$: (a) $n=0.46875$ incompressible phase (IP), (b) $n=0.40625$ cluster incompressible phase (CIP), (c) $n=0.3215$ grain-boundary superfluid (GBSF). (d) $\theta=\frac{3 \pi}{8}$, $\phi=\frac{\pi}{8}$, $\gamma=0.0532$, and  $L=20$. We fix $n$ to $0.25$ and observe coexistence of vacuum and superfluid regions. (e-f) diagonal stripe solids (DiagSTRS) at $\theta=\frac{\pi}{4}$, $\phi=\frac{\pi}{4}$, $\gamma=0.1854$, $L=20$: (e) $n=0.25$, (f) $n=0.5$. (g-h) $\theta=\frac{\pi}{4}$, $\phi=\frac{\pi}{4}$, and $L=40$: (g) grain-boundary superfluid (GBSF) at $n=0.321$ and $\gamma=0.166$, (h) incompressible phase (IP) at $n=0.325$ and $\gamma=0.169$. } 
\label{FIG3}
\end{figure*}

Figure~\ref{FIG1} shows the ground-state phase diagram at $\phi=\pi/8$, and  polar angles $\theta=\pi/16$ (a), $\pi/4$ (b), $3\pi/8$ (c).  At $\theta=\pi/16$ (Fig.~\ref{FIG1} (a)), the phase diagram features a SF, a CB stabilized at $n=0.5$, and CBSS phase.  We investigated the SF-CB transition ($\gamma\sim 0.055 $) at filling factor $n=0.5$ and did not find any evidence of hysteretic behavior in superfluid density $\rho_s$ and structure factor $S(\pi, \pi)$ as a function of the dipolar interaction strength $\gamma$. We used a step $\Delta \gamma=0.00065$. We were also unable to detect any supersolid phases. If either exists, it would be within a range narrower than $\Delta \gamma$.
Upon doping with particles or holes from half-filling, we enter the CBSS phase. Here, diagonal long-range order and off-diagonal long-range order coexist, that is, superfluid density $\rho_s$ and structure factor $S(\pi, \pi)$ are simultaneously finite. For large enough doping, on both particle and hole sides, the supersolid disappears via a second-order phase transition of Ising type in favor of a SF phase. We notice that this phase diagram is pretty much unchanged from the one at  $\phi=0$ (see Ref. \cite{PhysRevA.103.043333}). This is because at small polar angles, the dipolar contribution to the Hamiltonian parameters does not change significantly as a function of $\phi$, which can be seen in Fig.~\ref{fig:hamparamsvsthetaphi}. Notice that one may expect other solid phases stabilized at rational filling factors, e.g. a star solid at $n=1/4,3/4$ (see e.g.~\cite{Wu:2020kf}), for larger dipolar interaction strength (not explored here).

%%%%%%%%%%%%%%%%%%%%%%%%%%%%%%%%%%%%%%%%%%%%%%%%%%%%%%%%%%%%%

At $\phi=\pi/8$ and $\theta=\pi/4$, Fig.~\ref{FIG1}(b), the phase diagram features SF, StrS, StrSS, a cluster supersolid (CSS), a grain-boundary superfluid (GBSF), and an incompressible (IP) and cluster incompressible (CIP) phase. At this polar angle, the dipolar interaction along the $x$ axis is attractive stabilizing a stripe solid phase at filling factor $n=0.5$ and $\gamma\gtrsim 0.0767$. In the StrS, particles arrange themselves such that fully occupied horizontal rows alternate with emtpy ones.  We have studied the transition from SF to StrS at half filling and observed that a supersolid intervenes in between, for a narrow range $0.0748 \lesssim \gamma \lesssim0.0767$. For $0.0767\lesssim\gamma\lesssim0.0960\; ({0.1000})$, a StrSS phase (shaded pink area) also appears upon doping the stripe solid with holes (particles).   We notice that for larger $\gamma$ and large enough doping, spacing between stripes can be irregular (we will discuss this below in more details for the incompressible phase).  The StrSS disappears in favor of a SF via a second order transition of Ising type.

When $\gamma$ is further increased, upon doping the half-filling solid, the system stays incompressible. This incompressible phase (IP) first appears on the hole side. The IP corresponds to a succession of incompressible ground states with rational filling factors (this succession will become dense in the thermodynamic limit), similar to the classical devil's staircase~\cite{Hubbard:1978im,Fisher:1980be,Bak:1982ev}. In the IP (cyan shaded area), particles arrange themselves in stripes, similarly to the StrS at $n=0.5$, but with the difference that the spacing between stripes can be irregular to accommodate a specific filling. Figure~\ref{FIG3}(a) shows an example of a density map of the IP phase at $\gamma=0.1096$ and $n=0.46875$. Here, each circle corresponds to a single lattice site, and its radius is proportional to the local density. The transition IP-StrSS is of first order (marked in dark blue) as confirmed by a discontinuity in density and superfluid stiffness, and hysteretic behavior. At lower densities, we observe what appears to be a smooth changeover to a different type of IP (yellow shaded area) where particles arrange in horizontal clusters of length smaller than L. These clusters align along a direction that makes an angle of $7^\circ-10^\circ$ with the horizontal (depending on the density). We call this phase a cluster IP (CIP). In figure~\ref{FIG3}(b), we show an example of a density map of this cluster solid. This particle arrangement results from the competition between attractive interaction along the $x$-direction which favors a stripe solid structure, and attractive interaction along the positive diagonal which favors a diagonal solid structure. We notice that the size of the horizontal particle clusters  and their relative position vary slightly with density throughout the CIP region. Consequently, the value of $\mathbf{k}$  for which the structure factor peaks also varies.  

For $\gamma\gtrsim 0.106$ and upon decreasing density, the CIP disappears in favor of a grain-boundary superfluid (shaded light blue region). Here, extended solid regions of stripe solid at 1/3 filling, i.e. a single filled stripe alternating with two empty ones,  are separated by extended defects --grain boundaries-- which support superfluidity  along the direction of the boundary. In figure~\ref{FIG3}(c), we show an example of a density map of this phase at $\gamma=0.1096$ and $n=0.3215$. 
For the system size considered, we observe minor system size dependence of the one-dimensional superfluid stiffness across the GBSF-CIP boundary and do not further resolve it. 
Upon further decreasing the density, we enter a supersolid phase (shaded purple area in figure~\ref{FIG1}(b)) that we call cluster supersolid (CSS). Here, particles are also arranged in horizontal clusters much like in the cluster solid but the system also supports a superfluid response. We notice that similar phases have also been observed in lattice bosonic systems with soft-shoulder interactions~\cite{PhysRevLett.116.135303,PhysRevLett.123.045301}.  In going from  GBSF to cluster supersolid, within our resolution, we do not observe any obvious discontinuity in density or in the superfluid stiffness in the direction perpendicular to the boundary neither we observe hysteretic behavior which would all signal a first-order phase transition. Rather, these observables behave smoothly. We also do not observe considerable finite size dependence in our results for the system size considered and therefore we do not fully resolve the nature of this phase boundary. Upon increasing doping, CSS disappears in favor of a SF. Here, one would expect a second order phase transition. Nonetheless, since the structure factor peaks at slightly different $\mathbf{k}$-vectors for different densities, doing finite size scaling becomes complicated.
For $\gamma\lesssim 0.106$, the GBSF no longer intervenes between CSS and CIP  rather, for $L>20$, we find a very narrow region of metastability (green thick line) separating CSS from CIP. Finally, across the CSS-StrSS boundary, we observe a smooth changeover from stripe to cluster structure. 

We notice that we have found an extended metastable (MS) region (shaded green area in figure~\ref{FIG1}(b)) for $0.096 \lesssim\gamma\lesssim 0.0985$, where, given a certain $\gamma$ and $n$, we find the system in either a cluster supersolid, a stripe supersolid or an incompressible phase depending on the initial conditions of the simulation. The metastability is likely due to the onset, in this region, of competition between different density-density orders (cluster vs. stripe) and the competition between compressible (supersolids) vs incompressible phases away from half-filling. Indeed, four different phases are stabilized around this region. 

On the particle side of the stripe solid the situation is much simpler. The IP phase disappears in favor of the StrSS via a first order phase transition (thick dark blue line) as indicated by a jump in density and superfluid density as a function of chemical potential. 

Finally,  we  notice that overall the superfluid response is anisotropic, with the superfluid stiffness along the $x$-direction being larger than the one along the $y$-direction.

In Fig.~\ref{FIG1}(c), we plot the phase diagram at $\phi=\pi/8$ and $\theta=3\pi/8$ which, at filling factor $n<1.0$, only features a SF phase. Due to the competition between attractive and repulsive parts of the off-site interaction, no stable solid is observed in the parameter regime considered. The SF phase disappears via a first order phase transition in favor of the vacuum, as marked by the shaded dark blue region. Here, one observes coexistence of SF and vacuum. This is shown in Fig.~\ref{FIG3}(d), where, at fixed density $n=0.25$, one observes compressible, superfluid stripes of particles arranged at an angle and separated by regions of `vacuum'. Here, by vacuum, we refer to the regions of the lattice where the average density is either zero or much smaller than the density in the SF regions. We would expect solid and supersolid phases to be stabilized for $n>1$.

\begin{figure*}[th]
\centering
\includegraphics[trim=0.5cm 0cm 0cm 0cm, clip=true, width=1.0\textwidth]{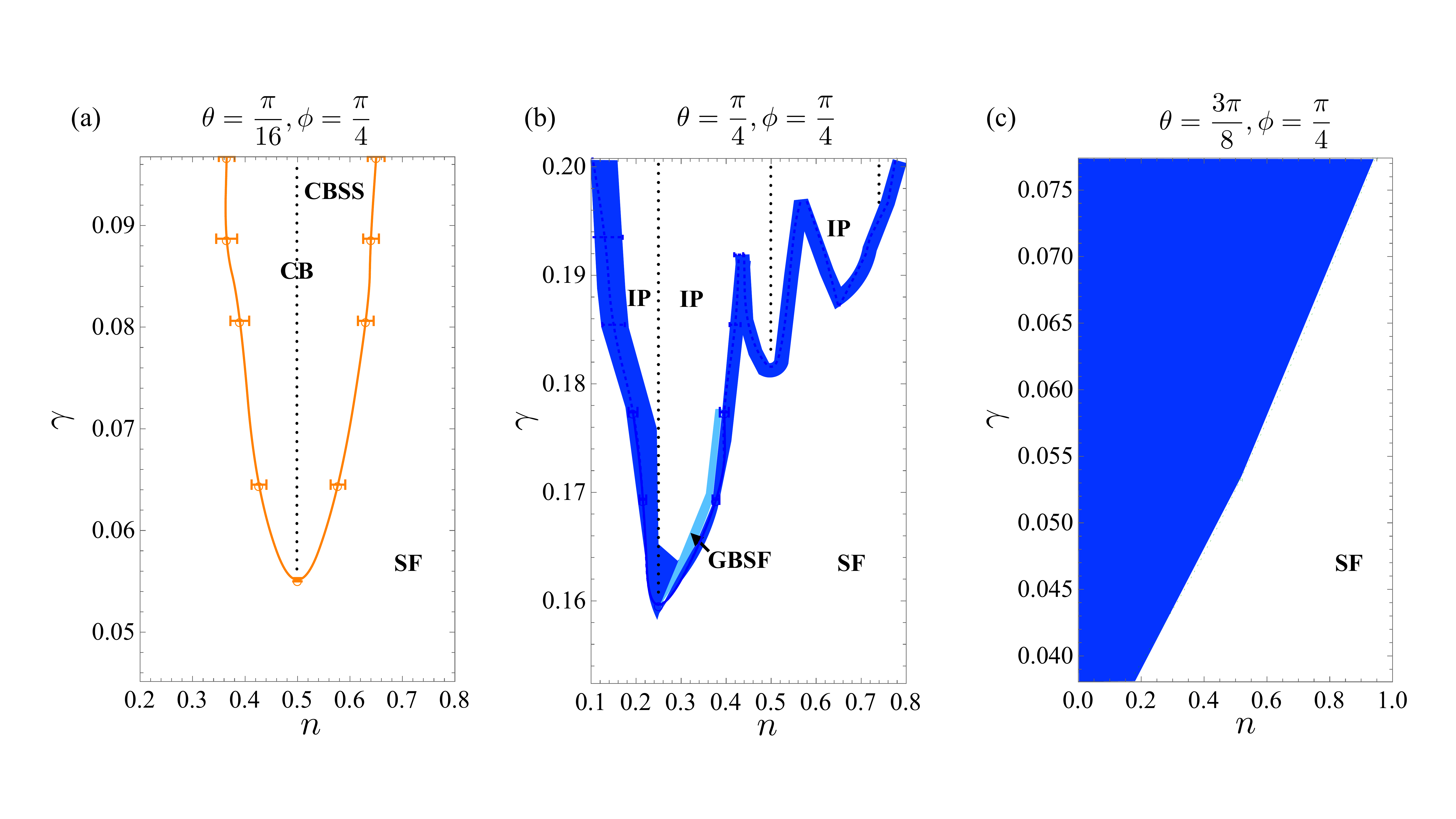}
\caption{Ground-state phase diagrams for $\phi=\frac{\pi}{4}$, $\theta=\frac{\pi}{16}$(a), $\theta=\frac{\pi}{4}$(b), and $\theta=\frac{3\pi}{8}$(c). The $x$-axis is the filling factor $n$ and the $y$-axis is the dipolar interaction strength $\gamma$. (a) checkerboard solid (CB) at $n=0.5$ dotted line, superfluid (SF), and checkerboard supersolid (CBSS); (b) diagonal stripe solid (DiagStrS) at $n=\frac{1}{4}$, $\frac{1}{2}$, and $\frac{3}{4}$ (dotted lines), incompressible ground states (IP) stabilized at rational filling factors, grain-boundary superfluid (GBSF), superfluid phase (SF). Dark blue regions in (b) and (c) represent first-order phase transitions.} 
\label{FIG2}
\end{figure*}

Figure~\ref{FIG2} shows the ground-state phase diagrams at $\phi=\pi/4$ and polar angles $\theta=\pi/16$ (a), $\pi/4$ (b), $3\pi/8$ (c). At $\theta=\pi/16$ (Fig.~\ref{FIG2}(a)), the phase diagram features a SF, a CB stabilized at $n=0.5$, and CBSS phase. As expected for small polar angles, this phase diagram is very similar to Fig.~\ref{FIG1}(a) and Fig. 3(b) in Ref.~\cite{PhysRevA.103.043333}. As before, we were unable to resolve any hysteretic behavior or a supersolid phase at half-filling with a step size $\Delta \gamma=0.00065$.

%%%%%%%%%%%%%%%%%%%%%%%%%%%%%%%%%%%%%%%%%%%%%%%%%%%%%%%%%%%%%
At $\phi=\pi/4$ and $\theta=\pi/4$ (Fig.~\ref{FIG2}(b)), we observe significant qualitative changes in the phase diagram compared to Fig.~\ref{FIG1}(b). At this azimuthal angle, the interaction along the positive diagonal is attractive while the interactions between nearest neighbors and the interaction along the negative diagonal are all repulsive. As a result, the model stabilizes a variety of incompressible phases with particles arranged along the positive diagonal. The diagonal stripe solid (DiagStrS) at $n=1/4$ for $\gamma\gtrsim 0.159$ and at $n=1/2$ for $\gamma\gtrsim 0.1815$ are shown in Fig.~\ref{FIG3}(e) and (f) for $L=20$. We notice that, at $n=1/2$, we observe a diagonal solid with two consecutive filled diagonals followed by two consecutive empty diagonals. This is because the repulsion along the negative diagonal is stronger than the one along $x$- and $y$-direction. There also exists a solid phase at $n=3/4$ and $\gamma>0.196$ with three consecutive filled diagonals followed by one empty diagonal. We investigated the SF-DiagStrS transition at fixed filling factor, and found hysteresis curves as a function of the interaction strength $\gamma$ for the superfluid density $\rho_s$ and structure factor $S(\pi/2, -\pi/2)$,
signalling a first-order phase transition. 
On the hole side of the quarter filling DiagStrS, for $0.159 \lesssim \gamma \lesssim 0.176$ , we find that the solid phase disappears in favor of a SF via a first-order phase transition (dark blue shaded area) as clearly indicated by a jump in the density and in the superfluid density (not shown here). For larger $\gamma$, instead, an incompressible phase intervenes between the DiagStrS and SF. The nature of the IP phase is the same as what discussed above but with particles arranged on diagonals, similarly to the quarter filling case, and  spacing between filled diagonals which can be irregular. On the particle side, things are a bit more complex. At lower $\gamma$ values ($0.159 \lesssim\gamma\lesssim 0.164$), in the proximity of the onset of the DiagStrS at quarter filling, we observe a first-order DiagStrS-GBSF phase transition. This grain-boundary superfluid (light blue shaded region) is similar to the one discussed above. Here, regions of DiagStrS at 1/3 filling (one filled diagonal followed by two empty ones) are separated by extended defects --grain boundaries-- which support a superfluid response. Figure~\ref{FIG3}(g) shows a density map of this phase where we see that regions of DiagStrS at filling 1/3 are separated by superfluid grain-boundaries.  For larger $\gamma$, upon doping the quarter filling solid, we enter the IP phase. Here, particles are arranged on filled diagonals which can be not uniformly spaced. For large enough doping, the IP phase disappears in favor of the GBSF  which eventually disappears in favor of a SF via a first-order phase transition. The GBSF phase disappears altogether for $\gamma\gtrsim 0.178$. At larger $\gamma$, the IP occupies a large region in the parameter space. An example of the IP phase at filling $n=0.325$ is shown in Fig.~\ref{FIG3}(h), where we observe filled diagonal with some unequal spacing between them. 

%%%%%%%%%%%%%%%%%%%%%%%%%%%%%%%%%%%%%%%%%%%%%%%%%%%%%%%%%%%%%

At $\phi=\pi/4$ and $\theta=3\pi/8$ (Fig.~\ref{FIG2}(c)), the phase diagram looks much like the one in Fig.~\ref{FIG1} (c). Again, the competition between attractive and repulsive part of the interaction inhibits solid formation in the parameter range considered.

\begin{figure}[h]
\centering
\includegraphics[trim=5.cm 3cm 0cm 4cm, clip=true, width=0.5\textwidth]{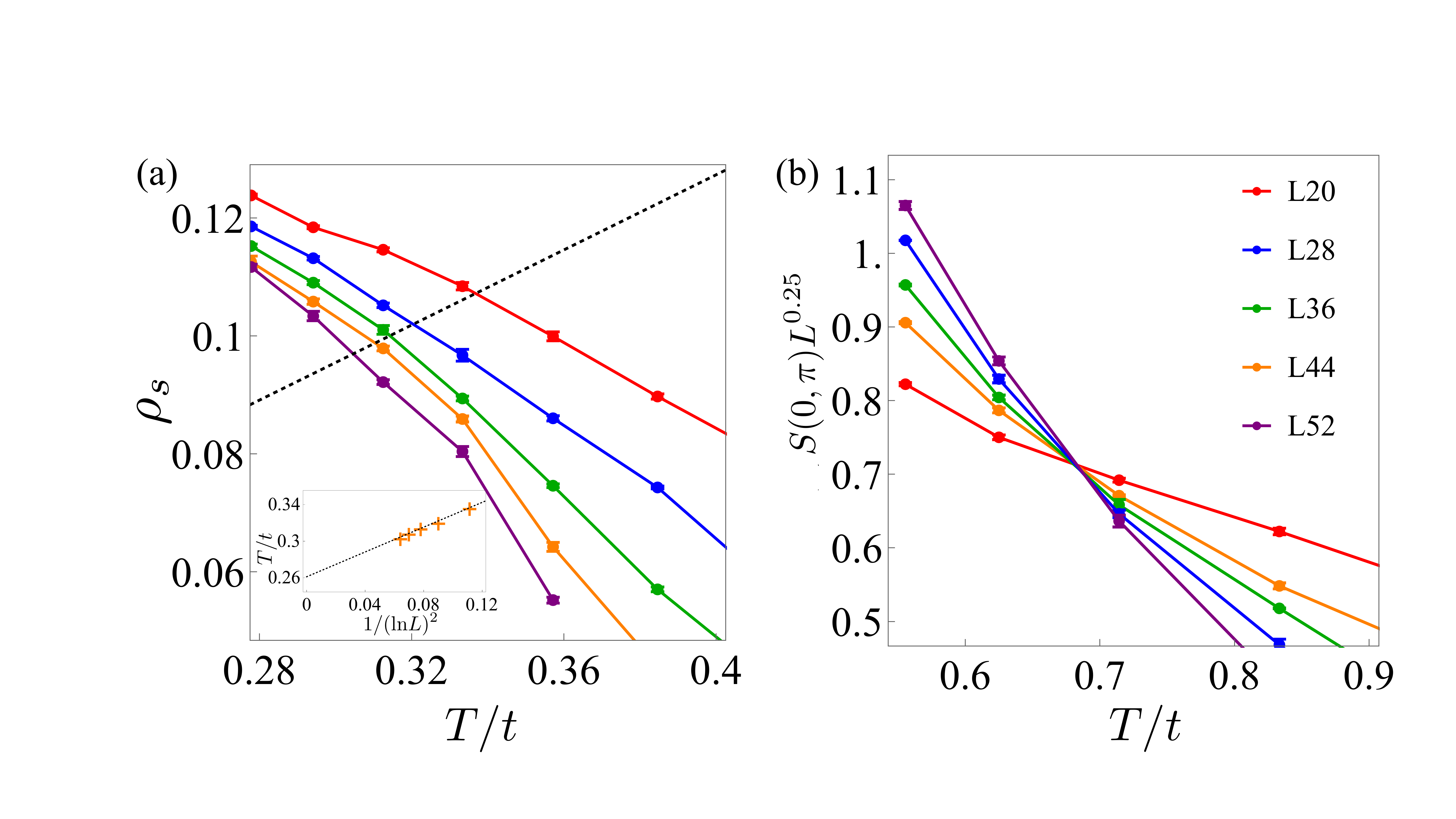}
\caption{Parameters $\theta=\frac{\pi}{4}$, $\phi=\frac{\pi}{8}$, $\gamma=0.0968$, and $n=0.5675$. Upon increasing the temperature, thermal fluctuations destroy the stripe supersolid phase in favor of a normal fluid in two steps. First, superfluidity is destroyed and the stripe supersolid becomes a stripe solid via a Kosterlitz–Thouless phase transition. Then, the stripe solid phase melts into a normal fluid via a two-dimensional Ising transition. In (a) we show $\rho_s$ as a function of $T/t$ for $L=20$ (red), 28 (blue), 36 (green), 44 (orange), and $52$ (purple). The dashed line is $T/t\pi$. Inset: intersection points between the $T/t\pi$ line and the $\rho_s$ versus $T/t$ curves for each $L$ are used to extract $T_c/t\sim0.26 \pm 0.02$. (b) scaled structure factor with $2\beta/\nu=0.25$ for $L=28, 36, 44, 52$. The crossing determines the critical temperature $T_c/t = 0.68\pm0.01$.}
\label{FIG4}
\end{figure}

Finally, we briefly discuss the robustness of the stripe supersolid against thermal fluctuations. We find the solid order to be the most robust against thermal fluctuations. We find that superfluidity in the stripe supersolid phase disappear first via a Kosterlitz-Thouless transition~\cite{Kosterlitz:1973fc} while the diagonal order survives at larger temperatures (see also Ref.~\cite{PhysRevA.100.063614}).
In Fig.~\ref{FIG4}(a), we show the superfluid density $\rho_s$ as a function of $T/t$ for $L=20$, 28, 36, 44, and 52 at $\theta=\pi/4$, $\phi=\pi/8$, $\gamma=0.0968$, and $n=0.5675$. In the thermodynamic limit, a universal jump is observed at the critical temperature given by $\rho_s(T_c)=2m k_B T_c/\pi \hbar^{2}$.  Here, $m$ is the effective mass in the lattice, $m=\hbar^2/2ta^2$. In a finite size system this jump is smeared out as shown in Fig.~\ref{FIG4}(a). To extract the critical temperature in the thermodynamic limit, we apply finite-size scaling to $T_c(L)$. From renormalization-group analysis  one finds $T_c(L)=T_c(\infty)+\frac{c}{\ln^2(L)}$, where $c$ is a constant and $T_c(L)$ is determined from $\rho_s(T_c,L)=2m k_B T_c/\pi \hbar^{2}$~\cite{PhysRevLett.39.1201,Kosterlitz_1974,Ceperley:1989hb}.
The dashed line in Fig.~\ref{FIG4}(a) corresponds to $\rho_s=T/t\pi$ ( $\hbar=1$, $k_B=1$, lattice step $a=1$ )  and its intersection points with each $\rho_s$ vs. $T/t$ curve are used to find $T_c$ as shown in the inset. We find $T_c/t=0.26 \pm 0.02$. Above this temperature the system is in a StrS. The solid order melts in favor of a normal fluid via a two-dimensional Ising transition.  We use standard finite size scaling as shown in Fig.~\ref{FIG4}(b), where we plot the scaled structure factor $S(0,\pi)L^{2\beta/\nu}$, with $2\beta/\nu=0.25$  as a function of $T/t$ for $L=20$, 28, 36, 44, 52. The crossing indicates a critical temperature $T_c/t = 0.68\pm0.01$.

\section{Experimental Realization}
\label{sec:sec5}

Dipolar systems can be created using atoms with magnetic dipole moments such as Cr
\cite{Griesmaier:2005fd,Naylor:2015bs}, Er \cite{Aikawa:2012ic,Aikawa:2014if,Baier:2016ga}, and Dy \cite{Lu:2012bd,Lu:2011hl}, polar molecules such as Er$_{2}$ \cite{Frisch:2015gm}, KRb \cite{Yan:2013fna}, NaK \cite{Seesselberg:2018ff}, and Rydberg dressing techniques \cite{Schauss:2015ch}. To create a two-dimensional system, the laser along $z$ direction should be able to be adjusted independently. By appropriately choosing intensity, wavelength, and cross angle between the two beams along the $z$ direction, a desired two-dimensional lattice system can be generated. With implementing three pairs of magnetic coils or metal plates along three perpendicular directions, the strength and direction of the dipoles in the lattice can be adjusted in three-dimensional space. While the tunneling amplitude can be controlled by changing light intensity, the on-site interaction can be adjusted using Feshbach resonance. Both the filling factor $n$ and temperature $T$ can be adjusted by manipulating the evaporation before loading into lattices. In general, a deeper evaporation gives a lower temperature with less atoms left. For atomic species with atomic mass around 150 amu loaded in optical lattices formed by 532 nm lasers, both critical temperatures in FIG.~\ref{FIG4} for the vanishing of the superfluidity and spatial structure are around few nano-kelvins.

The parameter $\gamma$ depends on the dipole moments and the masses of atomic species. It also depends on the lattice spacing. For magnetic atoms, the magnetic dipole moments are not large enough to observe quantum phases other than superfluidity. Indeed, in an optical lattice with a lattice constant $\sim$ 266 nm one has $\gamma \sim 0.0018$ for Cr,  $\gamma \sim 0.008$ for Er, $\gamma \sim 0.016$ for Dy. For Er$_{2}$ one gets $\gamma \sim 0.06$ which is still too small to realize solid or supersolid phases in the setup studied. Nonetheless, likely, in a bilayer geometry, solid orders could be observed with Er$_{2}$ molecules~\cite{Grimmer:2014kx}. 
While the preparation of and subsequent observation with polar molecules are more challenging, these systems are better  candidates to explore the quantum phases discussed here. Polar molecules possess dipole moments around one to few Debye, depending on different quantum number states corresponding to, e.g., $\gamma \sim 0.8$ for KRb, $\gamma \sim 4$ for RbCs. Another way to adjust $\gamma$ is to continuously change the lattice spacing in two dimensions, this topic is still challenging up to today. It is recommended to refer to \cite{phelps2019dipolar} for some pioneering works.

To observe a supersolid state is to confirm the existence of both crystalline order and global phase coherence in a system at the same time. Quantum gas microscopes, which can give a single-site-resolved resolution \cite{sherson2010, simon2011,yang2021siteresolved}, are capable for observing atom number density distribution with periodic patterns, which include CB, CBSS, DiagSS phases and so on. The global phase coherence can be observed using time-of-flight observation after releasing atoms from the lattice \cite{Greiner:2002es, RevModPhys.80.885}. Thus all phases mentioned above can be well captured using a combination of these two methods.

\section{Conclusion}
In this work, we have studied a system of dipolar bosons in a square optical lattice. Dipole moments are parallel to each other and their direction can be fixed in the three-dimensional space. The effective model describing the system is the extended Bose-Hubbard model. We start from the parameters that can be experimentally tuned, e.g. scattering length, dipolar interaction strength, optical lattice depth, and we calculate  parameters entering the effective model. Overall, besides superfluidity,  we have found a variety of solid and supersolid phases, e.g.  checkerboard and stripe solids and supersolids, depending on the direction of the dipoles. For  angles $\theta=\frac{\pi}{4}$ and $\phi=\frac{\pi}{8}$, we have observed a very rich phase diagram which includes a cluster incompressible phase, a cluster supersolid phase, a metastable phase, and a grain-boundary superfluid phase. In both the cluster incompressible phase and the cluster supersolid phase, particles form horizontal clusters. These clusters order themselves along a direction at an angle with the horizontal. In the grain-boundary superfluid, regions of solid order are separated by extended defects that support superfluidity. We have also briefly discussed the robustness of the stripe supersolid against thermal fluctuations. All phases can in principle be accessible within ultracold experiments with polar molecules.  In the future, a thorough finite-temperature study of this system can pinpoint where, in the parameter space, higher critical temperatures exist and therefore provide further guidance to experiments.
\label{sec:sec6}

{\textit{Acknowledgements}}  
Chao Zhang would like to thank Jie Wang for fruitful discussions. The computing for this project was performed at the OU Supercomputing Center for Education $\&$ Research (OSCER) at the University of Oklahoma (OU) and the cluster at Clark University.

\bibliography{induce}

\end{document}